\documentclass[a4paper,useAMS]{iau-JDSS}
\usepackage{color,graphicx}
\usepackage[english]{babel}
\usepackage[colon]{natbib}

\title[The Near Earth Asteroid associations] 
{The Near Earth Asteroid associations}

\author[T.J. Jopek]   
{Tadeusz J. Jopek$^1$,  %
 }
\affiliation{$^1$Astronomical Observatory Institute, Faculty of Physics, A.M. University, Poznan, Poland 
\\ [\affilskip]
}
\pubyear{2013}
\volume{Volume 16}  
\pagerange{1-2}
\date{?? and in revised form ??}
\jname{Highlights of Astronomy, Volume 16}
\editors{Montmerle T. prime ed.}
\begin{document}
\maketitle
\vspace{-6pt}
\begin{abstract}
We have made an extensive search for grouping  amongst  the near Earth asteroids (NEAs). We used two D- functions and  rigorous cluster analysis approach. 
We have found several new groups (associations) among the NEAs:  the objects moving on similar orbits with small minimum orbital intersection distances (MOID) with the Earth trajectory.
Reliability of some of these groups is quite high. 
\vspace{-6pt}
\end{abstract}
\vspace{-12pt}
\section{Introduction} 
The existence of the main belt asteroid families is beyond the doubts. However existence of groups  among the NEAs,  with members of common origin, suggested by similarity between their orbital  parameters \citep{Drummond1991, Obrubov1991, Fu2005, Jopek2011, Schunova2012} is not yet generally accepted.  Many groups found by these authors might be attributed to chance alignments.

In this study we have searched for associations amongst $9004$ NEA's osculating orbits \citep{Neodys2012} by two strict cluster analysis methods, similar to that used in \cite{Jopek2011}. Two D-functions were used: $D_{SH}$ introduced in \cite{SouthworthHawkins1963}  and $D_H$ introduced in \cite{Jopek1993}. The orbital similarity thresholds corresponded to $1$\% probability that given group was found by chance alignment. The clusters were detected by a single neighbour linking technique.
\vspace{-12pt}
\section{Results and discussion}
Using two searching methods we found $20$ groups of ten or more members: $13$ groups with the $D_{SH}$ function and $9$ groups using $D_H$ function.  Among 20 formally separated groups we have selected 10 associations.  All but two has been found in both searches, however with different amount of members. Associations No 1 and 2 (SH -179 and SH-380)  were detected only with $D_{SH}$ function. Of course when the values of the similarity thresholds  has been slightly increased, these groups were found also with $D_H$ function. 
Associations No 3-7 were identified with both functions with considerable amount of members in common.
In case of associations 8,9 and 10 the results shown to be more complicated: several overlapping subgroups were detected for which the common members were less numerous. 
In Table \ref{meanelements} we see that for all groups the mean orbital inclinations are smaller than $10$ degrees. It is not unexpected, we should recall that all NEAs but two have $i<75$ degrees, and for $\sim 4500$ orbits the inclinations are smaller than $10$ degrees. 

\begin{table}
\caption{The list of $10$ NEAs associations of $10$ or more members found in this study. The codes of the associations include the distance function tag and the ordinal number of the asteroid from the NEODyS list which was identified as a first group member. The name of this asteroid is also given. $N$ is the amount of members in the group. In brackets the amount of the common members identified by both functions are given. Within each group the orbital elements were averaged by the method described in \cite{Jopek2006}. }  
\centering
\scriptsize
\begin{tabular}{|r|l|l|r|r|r|r|r|r|r|}
\hline
\multicolumn{1}{|c}{No}              &
\multicolumn{1}{|c} {NEA }             &
\multicolumn{1}{|c}{Code }           &
\multicolumn{1}{|c}{N}               &
\multicolumn{1}{|c}{ a [AU] }         &
\multicolumn{1}{|c }{ q [AU]}         &
\multicolumn{1}{|c| }{e }             &
\multicolumn{1}{|c| }{i [deg] }             &
\multicolumn{1}{|c| }{$\omega$ [deg] }             &
\multicolumn{1}{|c| }{$\Omega$ [deg] }            \\
\hline
  1 & '13553' Masaakikiyama &SH-179  &    53       &  2.133 &   1.154 &   0.459 &     5.7 &   157.3 &   143.5 \\
  2 & '89136' 2001US16      &SH-380  &    24       &  1.375 &   1.025 &   0.254 &     1.0 &    16.0 &   227.3 \\
  3 & '2368' Beltrovata     & SH-35  &    77(74)   &  2.128 &   1.221 &   0.426 &     2.7 &    32.2 &   301.6 \\
    &                       &  H-35  &   101       &  2.137 &   1.222 &   0.428 &     3.0 &    31.2 &   305.6 \\
  4 & '4660' Nereus         & SH-77  &    63(40)   &  1.864 &    .936 &   0.498 &     1.2 &    37.2 &    82.1 \\
    &                       & H-77   &    63       &  1.793 &   0.959 &   0.465 &     1.0 &    46.0 &    70.9 \\
  5 & '8014' 1990MF         & H-143  &    69       &  2.019 &   1.051 &   0.479 &     0.1 &   100.3 &   232.1 \\
    & '10860' 1995LE        & SH-167 &    65(61)   &  2.135 &   1.111 &   0.480 &     1.3 &   160.9 &   167.4 \\
  6 & '11054' 1991FA        & H-168  &    68       &  1.513 &   0.984 &   0.350 &     0.9 &    84.7 &   349.1 \\
    & '190491' 2000FJ10     &SH-866  &    25(23)   &  1.427 &   1.014 &   0.289 &     2.6 &   184.9 &   249.8 \\
  7 & '36017' 1999ND43      &SH-255  &    39(21)   &  1.425 &   1.012 &   0.290 &     2.6 &   148.5 &   227.1 \\
    & '256004' 2006UP       & H-1050 &    35       &  1.421 &   1.013 &   0.287 &     2.8 &   157.4 &   217.0 \\
  8 & '54509' YORP          & H-295  &    78       &  1.045 &   1.031 &   0.013 &     1.0 &   321.6 &   224.2 \\
    & '65717' 1993BX3       & SH-306 &    89       &  1.162 &   0.957 &   0.176 &     0.6 &   235.4 &   210.2 \\
    & '209215' 2003WP25     & SH-910 &    23       &  1.082 &   0.926 &   0.144 &     0.8 &    95.9 &   188.8 \\
  9 & '19356' 1997GH3       &  H-202 &    73       &  1.815 &   1.007 &   0.445 &     0.9 &    14.5 &   145.1 \\
    & '27002' 1998DV9       & SH-236 &    27(21)   &  1.974 &   1.015 &   0.486 &     3.3 &    10.4 &   130.3 \\
    & '1994GV'              & SH-1385&    85       &  1.676 &   1.015 &   0.394 &     0.1 &   296.2 &   243.3 \\
    & '2003DW10'            & H-2582 &    16       &  1.587 &   0.998 &   0.371 &     0.2 &   127.5 &    74.8 \\
 10 & '136564' 1977VA       & SH-423 &    17(1)    &  2.172 &   1.141 &   0.475 &     4.2 &   204.1 &   199.3 \\
    & '2061' Anza           &   H-23 &   181       &  2.202 &   1.046 &   0.525 &     0.2 &   201.2 &   182.1 \\
  & '4015' Wilson-Harrington& SH-60  &   196(134)  &  2.150 &   1.046 &   0.514 &     0.3 &   169.0 &   223.1 \\
\hline
\end{tabular}
\normalsize
\label{meanelements}
\end{table}
We have continued our study gradually decreasing thresholds. As was expected, amount of members of the group was decreasing; some groups were splitting and finally disappearing. Associations SH-306, SH-1385, H23 have proved to be the most robust one. 
On Figure 1, just for an example we have plotted the orbits of the members of association SH380.
\begin{figure}
\centerline{\includegraphics[width=4.0cm]{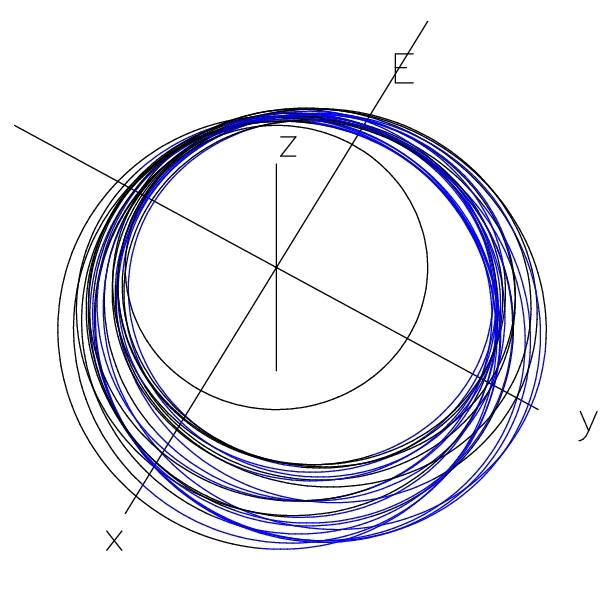}}
\footnotesize
\caption{
Association SH-380 plotted on the ecliptic plane. The orbits remarkably resemble a meteoroid stream. Association includes 24 NEAs: 89136, 26430 8,1994CJ1, 2003CC, 2003GA, 2004KG17, 2005HB4, 2005JT1, 2006HX30, 2006KL103, 2007HB15, 2008GR3, 2008LE, 2009DC12, 2009HG, 2009HH21, 2010CE55, 2011GR59, 2011GV9, 2011OK45, 2011OR5, 2011PU1, 2012GP1, 2012KT12.  The Earth circular trajectory is seen inside the association. }
\normalsize
\end{figure}

In this study we have shown, that using similar rigorous method as for the meteoroid orbits, the NEA's associations can be found easily. However, to ensure about their common origin, the long term numerical integration is needed.

\end{document}